\documentclass[aps,pre,amsmath,amssymb,twocolumn,groupedaddress,showpacs,eqsecnum]{revtex4}
\usepackage{graphicx}
\usepackage[utf8]{inputenc}
\usepackage[caption = false]{subfig}
\usepackage{float}
\usepackage{amsmath,amsfonts,amsthm,bm}
\usepackage{braket}

\usepackage{epsf}
\usepackage[toc]{appendix}
\newcommand{\be}{\begin{equation}}
\newcommand{\ee}{\end{equation}}
\newcommand{\ba}{\begin{eqnarray}}
\newcommand{\ea}{\end{eqnarray}}
\newcommand{\baa}{\begin{eqnarray}}
\newcommand{\eaa}{\end{eqnarray}}
\newcommand{\ed}{\end{document}}

\renewcommand{\baselinestretch}{1.2}
\date{\today}

\begin{document}
\title{Fast forward of adiabatic spin dynamics of entangled states}
\author{ Iwan Setiawan$^{1,2}$, Bobby Eka Gunara$^{1}$, Shumpei Masuda$^{3}$, and Katsuhiro Nakamura$^{4,5}$}
\affiliation{$^{(1)}$Department of Physics, Institut Teknologi Bandung, Jalan Ganesha 10, Bandung 40132, Indonesia\\
$^{(2)}$Department of Physics Education, University of Bengkulu, Kandang Limun, Bengkulu 38371, Indonesia\\
$^{(3)}$College of Liberal Arts and Sciences, Tokyo Medical and Dental University, Ichikawa, Chiba 272-0827, Japan \\
$^{(4)}$Faculty of Physics, National University of Uzbekistan, Vuzgorodok, Tashkent 100174, Uzbekistan\\
$^{(5)}$Department of Applied Physics, Osaka City University, Sumiyoshi-ku, Osaka 558-8585, Japan\\
}
\begin{abstract}
We develop a scheme of fast forward of adiabatic spin dynamics of quantum entangled states. We settle the quasi-adiabatic dynamics by adding the regularization terms to the original Hamiltonian and then accelerate it with use of a large time-scaling factor.  Assuming the experimentally-realizable candidate Hamiltonian consisting of the exchange interactions and magnetic field, we solved the regularization terms. These terms multiplied by the velocity function give rise to the state-dependent counter-diabatic terms.
The scheme needs neither knowledge of full spectral properties of the system nor solving the initial and boundary value problem. Our fast forward Hamiltonian generates a variety of state-dependent counter-diabatic terms for each of adiabatic states, which can include the state-independent one. We highlight this fact by using minimum (two-spin) models for a simple transverse Ising model,  quantum annealing and generation of entanglement.
\end{abstract}
\maketitle

\section{Introduction}
A shorter time in manufacturing products (e.g., electronics, automotives, plants, etc.) is becoming an important factor in nanotechnology. If we try to fabricate massive amount of such nanoscale structure, we should shorten the dynamics of each atom or molecule to get its desired target states in shorter time. In designing quantum computers, the coherence of systems is degraded by their interaction with the environment, and therefore the acceleration of adiabatic quantum dynamics is highly desirable. A theory to accelerate  quantum dynamics is proposed by Masuda and Nakamura \cite{1} with use of additional phase and driving potential. This theory aims to accelerate a known quantum evolution and to obtain the desired target state on shorter time scale, by fast forwarding the standard quantum dynamics. The theory of fast-forward can be developed to accelerate the adiabatic quantum dynamics \cite{2,3,4}, and constitutes one of the promising means to the shortcut to adiabaticity (STA) \cite{5,6,7,8,9,10,11}. The relationship between the fast-forward and the STA is nowadays clear \cite{4} (see also \cite{12,13}). The adiabaticity occurs when the external parameter of Hamiltonian is very-slowly changed. The quantum adiabatic theorem \cite{14,15,16,17,18} states that, if the system initially in an eigenstate of the instantaneous Hamiltonian, it remains so during the adiabatic process. Although the theory of fast forward of adiabatic quantum dynamics has been well developed for orbital dynamics, the corresponding study on quantum spin system remains in an elementary level \cite{19}. The scheme of fast forward of adiabatic spin dynamics will be important when the number of spins is plural and the quantum entanglement \cite{Nil-Chu} is operative. 

In this paper we shall develop a scheme of fast forward of adiabatic spin dynamics of quantum entangled states. We apply the scheme to two-spin systems described by a simple transverse Ising model\cite{24}, a minimum model for quantum annealing\cite{20,21} and a model for generation of entanglement\cite{22,23}, all of which are extremely important in the context of quantum computers. In Section \ref{FFADspin} we shall construct the scheme of fast forward of adiabatic quantum spin dynamics and elucidate its relation with the method of transitionless quantum driving. In Section \ref{couple-spin}, we shall apply the fast forward scheme to several coupled (two-spin) systems, and obtain a variety of state-dependent counter-diabatic terms to guarantee the accelerated entanglement dynamics. Section \ref{concl} is devoted to summary and discussions. Appendices give some technical details.
\section{Fast-forward of adiabatic spin dynamics}\label{FFADspin}

Consider the Hamiltonian for the spin systems to be characterized by the slowly time-changing parameter $R(t)$ such as the exchange interaction, magnetic field, etc. Then we can study the eigenvalue problem for the time-independent Schr\"{o}dinger equation :
\begin{equation}\label{satu}
H_0(R) \begin{pmatrix} 
  C_1(R) \\  \vdots\\ 
  C_N(R)
 \end{pmatrix} = E(R)\begin{pmatrix} 
   C_1(R) \\  \vdots\\
   C_N(R)
  \end{pmatrix},
 \end{equation}
where $R(t)= R_0 + \epsilon t$ is the adiabatically-changing parameter with $\epsilon \ll 1$. In Eq.(\ref{satu}), 
the quantum number $n$ for each eigenvalue and eigenstate is suppressed for simplicity.
Let us assume 
\begin{equation}\label{psi0}
\Psi_0(R(t)) =  \begin{pmatrix} 
  C_1(R) \\  \vdots\\
  C_N(R)
 \end{pmatrix} e^{-\frac{i}{\hbar}\int_{0}^{t}E(R(t'))dt'} e^{i\xi(t)},
\end{equation}
to be a quasi-adiabatic state, i.e., adiabatically evolving state. $\xi$ is the adiabatic phase \cite{14,15,16} defined by
\begin{eqnarray}\label{adiabatic}
\xi(t)& = & i \int_{0}^{t} dt'\Big(C_1^*\frac{\partial C_1}{\partial t} +...+ C_N^* \frac{\partial C_N}{\partial t}\Big)\\\nonumber
& = & i \epsilon \int_{0}^{t} dt'\Big(C_1^*\frac{\partial C_1}{\partial R} +...+ C_N^* \frac{\partial C_N}{\partial R}\Big).
\end{eqnarray}
For non-adiabatic processes, $\Psi_0(R(t))$ in Eq.(\ref{psi0}) does not satisfy  the time-dependent Schr\"{o}dinger equation (TDSE) and in order to impose it as the solution of the TDSE, the Hamiltonian must be regularized as 
\begin{equation}\label{reg}
H_0^{reg}(R(t)) = H_0(R(t)) + \epsilon\mathcal{\tilde{H}}_n(R(t)).
 \end{equation}
Then TDSE becomes
\begin{equation}
i \hbar \frac{\partial}{\partial t} \Psi_0(R(t))=(H_0+\epsilon \mathcal{\tilde{H}}_n)\Psi_0(R(t)).
\end{equation}
Here  $\mathcal{\tilde{H}}_n$ is the $n$-th state-dependent regularization term \cite{2}.
Substituting $\Psi_0(R(t))$ in Eq.(\ref{psi0}) into the above TDSE, 
we obtain:
\begin{eqnarray}
i\hbar (\epsilon \partial_R\textbf{C}-\frac{i}{\hbar}E\textbf{C}&-&\epsilon (\textbf{C}^{\dagger}\partial_R \textbf{C})\textbf{C} )
=H_0\textbf{C} + \epsilon  \mathcal{\tilde{H}}_n\textbf{C},\nonumber\\
&&\qquad \textbf{C}\equiv 
\begin{pmatrix} 
  C_1(R) \\  \vdots\\
  C_N(R)
 \end{pmatrix} .
\end{eqnarray}
While the order of $O(\epsilon^0)$ in the above equality gives the adiabatic eigenvalue problem in Eq.(\ref{satu}), 
the order of $O(\epsilon^1)$
leads to
\begin{equation}\label{sum2}
\mathcal{\tilde{H}}_n\begin{pmatrix} 
  C_1(R) \\  \vdots\\
  C_N(R)
 \end{pmatrix}  = i \hbar \begin{pmatrix} 
   \frac{\partial C_1(R)}{\partial R} \\  \vdots\\
  \frac{\partial C_N(R)}{\partial R}
  \end{pmatrix} -i\hbar \Bigg(  \sum_{j=1}^{N} C_j^*\frac{\partial C_j}{\partial R}\Bigg )\begin{pmatrix} 
    C_1(R) \\  \vdots\\
    C_N(R)
\end{pmatrix},
\end{equation}
which is the core equation of the present paper.
 
The fast forward state is defined by

\begin{equation}\label{psiff}
\Psi_{FF}(t) = \begin{pmatrix} 
  C_1(R(\Lambda(t))) \\  \vdots\\
  C_N(R(\Lambda(t)))
 \end{pmatrix}
 e^{-\frac{i}{\hbar}\int_{0}^{t}E((R(\Lambda(t'))))dt'} e^{i\xi((R(\Lambda(t))))}
 \end{equation}
where $\Lambda(t)$ is an advanced time defined by
\begin{equation}\label{lamda}
\Lambda(t) =  \int_{0}^{t} \alpha (t')dt',
\end{equation}
with the standard time $t$. $\alpha(t)$ is a magnification time-scale factor given by $\alpha(0) =1$,  $\alpha(t) > 1$  $(0 < t < T_{FF})$ and $\alpha(t) =1$ $(t\geq T_{FF})$. We consider the fast forward dynamics which reproduces the target state $\Psi_0(T)$ in a shorter final time $T_{FF}$ defined by
\begin{equation}
T = \int_{0}^{T_{FF}}\alpha(t)dt.
\end{equation}
The explicit expression for $\alpha(t)$ in the fast-forward range ($0\leq t \leq T_{FF} $) is typically given by \cite{2} as :
\begin{equation}
\alpha(t) = \bar{\alpha}-(\bar{\alpha}-1) \cos\left(\frac{2 \pi}{T_{FF}}t\right),
\end{equation}
where $\bar{\alpha}$ is the mean value of $\alpha(t)$ and is given by $\bar{\alpha} = T/T_{FF}$.

We now take a strategy: a product of the mean value $\bar{\alpha}$ of an infinitely-large time-scaling factor $\alpha(t)$ and an infinitesimally-small growth rate $\epsilon$ in the quasi-adiabatic parameter should satisfy the constraint $\bar{\alpha} \cdot \epsilon$ = $finite$ in the asymptotic limit $\bar{\alpha} \rightarrow \infty$ and $\epsilon \rightarrow 0$. 
Then, by taking the time derivative of $\Psi_{FF}$ in Eq.(\ref{psiff}) and using Eqs.(\ref{satu}) and (\ref{sum2}), 
we find (see Appendix A for details)
\begin{eqnarray}\label{TDSE-FF}
i \hbar \frac{\partial \Psi_{FF}}{\partial t} &=& \left(H_0(R(\Lambda(t))) + v(t) \mathcal{\tilde{H}}_n(R(\Lambda(t)))
\right) \Psi_{FF}\nonumber\\
&\equiv&  H_{FF} \Psi_{FF}. 
\end{eqnarray}
Here $v(t)$ is a velocity function available from $\alpha(t)$ in the asymptotic limit:
\begin{eqnarray}\label{velo}
v(t) &=& \lim_{\epsilon\to 0, \alpha \to \infty} \epsilon \alpha(t) \\\nonumber
 & = & \bar{v}\left(1-\cos \frac{2 \pi}{T_{FF}}t\right),
\end{eqnarray}
where $\bar{v}=\lim_{\epsilon\to 0, \alpha \to \infty} \epsilon \bar{\alpha}(=finite)$ is the mean of $v(t)$.

Consequently, for $0  \le t \le T_{FF}$,
\begin{eqnarray}\label{lambda2}
R(\Lambda(t))&=&R_0+\lim_{\epsilon\rightarrow 0, \bar{\alpha}\rightarrow \infty}\varepsilon\Lambda(t)\nonumber\\
&=&R_{0}+\int^{t}_{0}v(t')dt'\nonumber\\
&=& R_{0}+\bar{v}\left[t-\frac{T_{FF}}{2\pi}\sin\left(\frac{2\pi}{T_{FF}}t\right)\right].
\end{eqnarray}
In Eq.(\ref{TDSE-FF}), $H_{FF}$ is the driving Hamiltonian and $\mathcal{\tilde{H}}_n$ is the regularization term obtained from Eq.(\ref{sum2}) to generate the fast-forward scheme in spin systems. 

There is a relation between $\mathcal{\tilde{H}}_n$ in Eq.(\ref{sum2}) and Demirplak-Rice-Berry's  counter-diabatic term $\mathcal{H}$ \cite{5,6,7}. 
If there is an $n$-independent regularization term $\mathcal{\tilde{H}}$ among $\{\mathcal{\tilde{H}}_n\}$, we define $\mathcal{H} \equiv v(t) \mathcal{\tilde{H}}(R(\Lambda(t))) $ 
with use of $v(t) = \frac{\partial R(\Lambda(t))}{\partial t}$. Then Eq.(\ref{sum2}) becomes
\begin{equation}
\mathcal{H} \textbf{C} = i \hbar \partial_t \textbf{C}- i \hbar (\textbf{C}^{\dagger}\partial_t \textbf{C})\textbf{C},
\end{equation}
which can be rewritten as
\begin{equation}\label{ber}
\mathcal{H} |n \rangle = i \hbar \frac{\partial}{\partial t} |n \rangle - i \hbar  |n \rangle  \langle n|\frac{\partial}{\partial t}|n \rangle,
\end{equation}
where $|n \rangle$ means the $n$-th eigenstate of the Hamiltonian in Eq.(\ref{satu}). Operating both side of Eq.(\ref{ber}) on $\langle n|$, and summing over $n$, we have
\begin{equation}
\mathcal{H} \sum_{n} |n \rangle \langle n| = i \hbar \sum_{n} \frac{\partial}{\partial t} |n \rangle\langle n| - i \hbar \sum_{n}  |n \rangle  \langle n|\frac{\partial}{\partial t}|n \rangle \langle n|.
\end{equation}
Noting the completeness condition for the eigenstates : $\sum_{n} |n \rangle \langle n| = 1$, we have
\begin{equation}\label{ric}
\mathcal{H} = i \hbar \sum_{n} \left(\frac{\partial}{\partial t} |n \rangle\langle n| -  |n \rangle  \langle n|\frac{\partial}{\partial t}|n \rangle \langle n|\right),
\end{equation}
which agrees with Demirplak-Rice-Berry's formula. 
Therefore $v(t) \mathcal{\tilde{H}}(R(\Lambda(t))) $ corresponds to the counter-diabatic term. 
Using this correspondence, one may call $v(t) \mathcal{\tilde{H}}_n(R(\Lambda(t))) $ as a state-dependent counter-diabatic term.
Hereafter we shall be concerned with the fast forward of adiabatic dynamics of one of the adiabatic states (e.g., the ground state),  and thereby the suffix $n$ in $ \mathcal{\tilde{H}}_n$  will be suppressed.

Note: Demirplak-Rice-Berry(DRB)'s counter-diabatic(CD) term is state-independent, and can also  be reproduced by
the inverse engineering \cite{XChen} based on the Lewis-Riesenfeld's invariant theory \cite{8}. Inspired by the works \cite{12,Mart} on a streamlined version of the fast-forward method,
Patra and Jarzynski (PJ) \cite{Patra}proposed a framework for constructing  the STA from the velocity and acceleration flow field which characterizes the adiabatic evolution, providing compact expressions for both CD term and fast-forward potentials. Since the flow field is uniquely defined using each adiabatic eigenstate, PJ generates only one state-dependent CD term, which is not equivalent to DRB's CD term, although the equivalence will be recovered if two kind of CD terms will be projected onto each of adiabatic states. By contrast, our formalism generates plural number of state-dependent CD terms for each adiabatic state, which can include a state-independent one.

Now we investigate a single spin system in our scheme, and show the fast forward of adiabatic dynamics in Landau-Zener (LZ) model \cite{25,26} described by the spin Hamiltonian,
\begin{equation}\label{LZmodel}
H_0(R(t)) = \frac{1}{2} \bm{\sigma}\cdot{\bf B} = \frac{1}{2}\begin{pmatrix}
R(t)&\Delta & \\
\Delta & -R(t)\\
\end{pmatrix} ,
\end{equation}
where $\Delta$ is a constant. Equation (\ref{LZmodel}) has the eigenvalues $\lambda_\pm$ = $\pm\frac{\sqrt{R^2+\Delta^2}}{2}$ and eigenstates:
\begin{equation}\label{LZPsi}
\Psi_0^{\pm} = \begin{pmatrix} 
   C_1^{\pm}\\
  C_2^{\pm}  \end{pmatrix}= \begin{pmatrix}   
  -\Delta/s_{\pm} & \\
  \frac{R\mp \sqrt{R^2+\Delta^2}}{s_{\pm}}  \end{pmatrix}, 
\end{equation}
where
\begin{equation}
s_{\pm} \equiv \left[2\sqrt{R^2+\Delta^2}\left(\sqrt{R^2+\Delta^2}\mp R \right)\right]^{1/2}.
\end{equation}
Now we choose one of the states with $\lambda_+$ and $\Psi_0^+$, and consider the adiabatic dynamics where $R = R_0 + \epsilon t$. The adiabatically evolving state is :
\begin{equation}\label{landau}
\Psi_0(t) =  \begin{pmatrix}
-\frac{\Delta}{s_+}& \\
\frac{R-\sqrt{R^2+\Delta^2}}{s_+} \\

\end{pmatrix} e^{-\frac{i}{\hbar}\int_{0}^{t}\frac{\sqrt{R^2+\Delta^2}}{2} dt'} e^{\xi(t)}.
\end{equation}
Noting that $\mathcal{\tilde{H}}_{i j}$ is traceless ($\mathcal{\tilde{H}}_{11}$ = - $\mathcal{\tilde{H}}_{22}$) and Hermitian ($\mathcal{\tilde{H}}_{21}^*$ =$\mathcal{\tilde{H}}_{12}$), Eq.(\ref{sum2}) constitutes a rank = 2 linear algebraic equation for two unknowns ($\mathcal{\tilde{H}}_{11}$ and $\mathcal{\tilde{H}}_{12}$). With use of
\begin{eqnarray}\label{landau1}
&&\frac{\partial C_1}{\partial R} = -\frac{1}{2\sqrt{2}}\frac{\Delta}{Q^{5/2}} (Q-R)^{\frac{1}{2}} \\\nonumber
&& \frac{\partial C_2}{\partial R} = \frac{1}{2\sqrt{2}} \frac{(Q-R)^{\frac{1}{2}}(Q+R)}{Q^{5/2}},
\end{eqnarray}
we can solve Eq.(\ref{sum2}) for $\mathcal{\tilde{H}}$ as :
\begin{eqnarray}\label{res}
\mathcal{\tilde{H}}_{11} &=& 0\\\nonumber
\mathcal{\tilde{H}}_{12}&=& i \frac{\hbar}{2}\frac{\Delta}{Q^2}
\end{eqnarray}
with
$Q \equiv \sqrt{R^2+\Delta^2}$ and $\xi=0$.
The state-dependent counter-diabatic term
and
the fast-forward Hamiltonian are written respectively as
\begin{equation}\label{FFLDB}
\mathcal{H} =v(t) \tilde{\mathcal{H}}=
 \begin{pmatrix} 
  0 & v(t) i \frac{\hbar}{2}\frac{\Delta}{Q^2} \\
   
  -v(t) i \frac{\hbar}{2}\frac{\Delta}{Q^2}& 0 
 \end{pmatrix} 
\end{equation}
and
\begin{equation}\label{FFLD}
H_{FF} =  \begin{pmatrix} 
   \frac{R(\Lambda(t))}{2} & \frac{\Delta}{2}+v(t) i \frac{\hbar}{2}\frac{\Delta}{Q^2} \\
   
  \frac{\Delta}{2}-v(t) i \frac{\hbar}{2}\frac{\Delta}{Q^2}& -\frac{R(\Lambda(t))}{2} 
 \end{pmatrix}.
\end{equation}

The fast forward state is obtained from Eq.(\ref{psiff}) as
\begin{eqnarray}
\Psi_{FF} = \begin{pmatrix} 
   C_1^+(\Lambda(t)) \\
  C_2^+ (\Lambda(t))  \end{pmatrix}e^{-\frac{i}{\hbar}\int_{0}^{t}\frac{\sqrt{R(\Lambda(t'))^2+\Delta^2}}{2}dt'}. \nonumber\\
\end{eqnarray}
The total driving magnetic field is written as
\begin{equation}\label{landau2}
\textbf{B}_{FF}(t) =  \begin{pmatrix}
\Delta   \\
-v(t) \hbar \frac{\Delta}{R(\Lambda(t))^2+ \Delta^2}   &\\
R(\Lambda(t))
\end{pmatrix}.
\end{equation}

Choosing another eigenstate $\Psi_0^-$ in Eq.(\ref{LZPsi}), we can reproduce the  regularization term  in Eq.(\ref{res}) and the counter-diabatic term in Eq.(\ref{FFLDB}), and therefore these terms are state-independent. 
By applying Demirplak-Rice-Berry formula in Eq.(\ref{ric}), on the other hand, one can obtain the counter-diabatic terms $\mathcal{H}$ which agrees with Eq.(\ref{FFLDB}). So long as we shall stay in single spin dynamics, therefore,  Eq.(\ref{sum2}) conveys no new information beyond Eq.(\ref{ric}) : Both equations lead to the identical result. The situation will be dramatically changed when we shall proceed to a system of coupled spins, which shows entanglement dynamics.

\section{Two spin systems}\label{couple-spin}\label{geneSec}
We shall generalize the scheme to two-spin systems which shows entanglement dynamics\cite{Nil-Chu,sta-ent}. Here, the number of independent equations in Eq.(\ref{sum2}) is less than that of the unknown $\{\mathcal{\tilde{H}}_{ij}\}$ ($1 \leq i, j \leq 4$). Some extra strategy should be introduced. We assume the experimentally-realizable form for the regularization term ($\mathcal{\tilde{H}}$) in Eq.(\ref{sum2}), which includes the diagonal-exchange interaction $\tilde{J}_1 = \tilde{J}_1(R(t)), \tilde{J}_2 = \tilde{J}_2(R(t)), \tilde{J}_3 = \tilde{J}_3(R(t))$, offdiagonal-exchange interaction $\tilde{W}_1 = \tilde{W}_1(R(t))$, $\tilde{W}_2 = \tilde{W}_2(R(t))$, $\tilde{W}_3 = \tilde{W}_3(R(t))$, and 3-component magnetic field $\mathbf{\tilde{B}} = \mathbf{\tilde{B}}(R(t))$. The candidate for regularization Hamiltonian $\mathcal{\tilde{H}}$ takes the following form :
\begin{widetext}
\begin{equation}\label{cand-reg}
\mathcal{\tilde{H}} =\tilde{J}_1 \sigma_1^x \sigma_2^x + \tilde{J}_2 \sigma_1^y \sigma_2^y +\tilde{J}_3 \sigma_1^z \sigma_2^z+\tilde{W}_1(\sigma_1^x \sigma_2^y +\sigma_1^y \sigma_2^x)+\tilde{W}_2( \sigma_1^y \sigma_2^z +  \sigma_1^z \sigma_2^y)+\tilde{W}_3(\sigma_1^z \sigma_2^x+ \sigma_1^x \sigma_2^z) +\frac{1}{2}(\bm{\sigma}_1+\bm{\sigma}_2)\cdot \mathbf{\tilde{B}},
\end{equation}
where $\sigma_1^{x,y,z}$ and  $\sigma_2^{x,y,z}$ represent Pauli matrices for two spins. 
The regularization Hamiltonian in Eq. (\ref{cand-reg}) shows  an expression widely accepted in the context of magnetic materials. 
In Eq.(\ref{cand-reg}) we suppressed products of of three or more Pauli matrices, which do not exist in magnetic systems.
Likewise we ignored the spin-independent term, which gives a deviation from Tr$( \mathcal{\tilde{H}})=0$ and is not essential in thermodynamic properties. The regularization Hamiltonian including these extra terms not acceptable in magnetic systems will not be investigated in the present paper.
Arranging the bases as $\Ket{\uparrow\uparrow} $, $\Ket{\uparrow\downarrow} $, $\Ket{\downarrow\uparrow}$, and $\Ket{\downarrow\downarrow}$, we obtain the matrix form: 
\begin{equation}\label{cdterm}
\mathcal{\tilde{H}} = \begin{pmatrix} 
  \tilde{J}_3+\tilde{B}_z & \frac{1}{2}(\tilde{B}_x-i\tilde{B}_y)-i\tilde{W}_2+\tilde{W}_3 & \frac{1}{2}(\tilde{B}_x-i\tilde{B}_y)-i\tilde{W}_2+\tilde{W}_3 & \tilde{J}_1-\tilde{J}_2-i2 \tilde{W}_1  \\
  \frac{1}{2}(\tilde{B}_x+i\tilde{B}_y)+i\tilde{W}_2+\tilde{W}_3 &  -\tilde{J}_3 & \tilde{J}_1+\tilde{J}_2 &\frac{1}{2}(\tilde{B}_x-i\tilde{B}_y)+i \tilde{W}_2-\tilde{W}_3 \\
 \frac{1}{2}(\tilde{B}_x+i\tilde{B}_y)+i\tilde{W}_2+\tilde{W}_3 & \tilde{J}_1+\tilde{J}_2 & -\tilde{J}_3 & \frac{1}{2}(\tilde{B}_x-i\tilde{B}_y)+i\tilde{W}_2-\tilde{W}_3 \\
  \tilde{J}_1-\tilde{J}_2+i2\tilde{W}_1 & \frac{1}{2}(\tilde{B}_x+i\tilde{B}_y)-i \tilde{W}_2-\tilde{W}_3 & \frac{1}{2}(\tilde{B}_x+i\tilde{B}_y)-i\tilde{W}_2-\tilde{W}_3 & \tilde{J}_3-\tilde{B}_z
 \end{pmatrix}.
\end{equation}
\end{widetext}
We see: $\tilde{B}_y, \tilde{W}_1$ and $\tilde{W}_2$ contribute to the imaginary part of the matrix $\mathcal{\tilde{H}}$, 
while $\tilde{J}_1, \tilde{J}_2, \tilde{J}_3, \tilde{W}_3, \tilde{B}_x$ and $\tilde{B}_z$  to its real part.
The explicit expression for $\mathcal{\tilde{H}}$ in Eq.(\ref{cdterm}) greatly reduces the number of unknown $\{\mathcal{\tilde{H}}_{ij}\}$ and helps us to solve Eq.(\ref{sum2}).
As two-spin systems, we shall investigate: (A) a simple transverse Ising model; (B) a minimum model for quantum annealing; (C) a model for generation of entanglement. 

\subsection{Simple transverse Ising model}\label{3top}
First of all, we study a simple Ising transverse-field model \cite{24} where our scheme reproduces the state-independent counter-diabatic terms obtained by the method of transitionless  quantum driving. The Hamiltonian is written as 
\begin{equation}
H_0 = J(R(t))\sigma_1^z \sigma_2^z - \frac{1}{2}(\sigma_1^x+ \sigma_2^x)B_x(R(t))
\end{equation}
By using this bases : $\Ket{\uparrow\uparrow} $, $\Ket{\uparrow\downarrow} $, $\Ket{\downarrow\uparrow}$, and $\Ket{\downarrow\downarrow}$, we have
\begin{equation}\label{Ap-matr}
H_0=\left(
\begin{array}{cccc}
 J & -\frac{B_x}{2} & -\frac{B_x}{2} & 0 \\
 -\frac{B_x}{2} & -J & 0 & -\frac{B_x}{2} \\
 -\frac{B_x}{2} & 0 & -J & -\frac{B_x}{2} \\
 0 & -\frac{B_x}{2} & -\frac{B_x}{2} & J \\
\end{array}
\right)
\end{equation}
where the eigenvalue : $-J$, $J$, $-\sqrt{J^2+B_x^2}$, and $\sqrt{J^2+B_x^2})$. The normalized eigenvector are respectively:\\
$\begin{pmatrix} 
  0  \\
  -\frac{1}{\sqrt{2}}   \\
 \frac{1}{\sqrt{2}}   \\
 0    
 \end{pmatrix}$, $\begin{pmatrix} 
   -\frac{1}{\sqrt{2}}  \\
  0   \\
  0   \\
   \frac{1}{\sqrt{2}}    
  \end{pmatrix}$, $\begin{pmatrix} 
   \frac{B_x}{2\sqrt{B_x^2+J^2+ J\sqrt{B_x^2+J^2}}}  \\
   \frac{\sqrt{B_x^2+J^2}+J}{2\sqrt{B_x^2+J^2+ J\sqrt{B_x^2+J^2}}}   \\
   \frac{\sqrt{B_x^2+J^2}+J}{2\sqrt{B_x^2+J^2+ J\sqrt{B_x^2+J^2}}}  \\
    \frac{B_x}{2\sqrt{B_x^2+J^2+ J\sqrt{B_x^2+J^2}}}    
   \end{pmatrix}$, and 
    $\begin{pmatrix} 
      \frac{B_x}{2\sqrt{B_x^2+J^2- J\sqrt{B_x^2+J^2}}}  \\
      \frac{-\sqrt{B_x^2+J^2}+J}{2\sqrt{B_x^2+J^2- J\sqrt{B_x^2+J^2}}}   \\
      \frac{-\sqrt{B_x^2+J^2}+J}{2\sqrt{B_x^2+J^2- J\sqrt{B_x^2+J^2}}}  \\
       \frac{B_x}{2\sqrt{B_x^2+J^2- J\sqrt{B_x^2+J^2}}}    
      \end{pmatrix}$.  \\
      
Let us focus on the ground state (the 3rd state with the lowest energy ($-\sqrt{J^2+B_x^2}$), where $C_1 = C_4$ , $C_2 = C_3$, and $C_1, C_2, C_3$ and $C_4$ are real. 
From $R$-derivative of the normalization, we see
\begin{equation}\label{bor}
\frac{\partial C_2}{\partial R} C_2 + \frac{\partial C_4}{\partial R} C_4 = 0,
\end{equation}
and then $\xi=0$. 

Due to the symmetry $C_1 = C_4$ and $C_2 = C_3$ and noting the real nature of $\{\tilde{J},\tilde{W}, \tilde{B}\}$, Eq.(\ref{sum2}) for the regularization terms reduces to
\begin{eqnarray}\label{aku2x2}
&& i\hbar \frac{\partial  C_4}{\partial R}  =   \tilde{\mathcal{A}}_{1}C_4 + \tilde{\mathcal{A}}_{2}  C_2 , \\\nonumber
&&
i\hbar \frac{\partial  C_2}{\partial R} =                \tilde{\mathcal{A}}_{3} C_4+  \tilde{\mathcal{A}}_{4}C_2,
\end{eqnarray}
where
$\tilde{\mathcal{A}}_{1}  = \tilde{\mathcal{H}}_{11}+\tilde{\mathcal{H}}_{14}=\tilde{J}_{1}-\tilde{J}_{2}+\tilde{J}_{3}$,
$\tilde{\mathcal{A}}_{2} =\tilde{\mathcal{H}}_{12}+ \tilde{\mathcal{H}}_{13}=\tilde{B}_{x}-2i\tilde{W}_{2}$, 
$\tilde{\mathcal{A}}_{3}  = \tilde{\mathcal{H}}_{21}+\tilde{\mathcal{H}}_{24}=\tilde{B}_{x}+2i\tilde{W}_{2}$, and 
$\tilde{\mathcal{A}}_{4}  = \tilde{\mathcal{H}}_{22}+\tilde{\mathcal{H}}_{23}=\tilde{J}_{1}+\tilde{J}_{2}-\tilde{J}_{3}$. 

To solve two-component simultaneous linear equations for ${\tilde{\mathcal{H}}_{ij}}$ in Eq.(\ref{aku2x2}), we should choose two independent real variables out of 5 real variables $(\tilde{J}_{1}, \tilde{J}_{2}, \tilde{J}_{3}, \tilde{W}_{2}, \tilde{B}_{x} )$ appearing in $\{\tilde{\mathcal{A}}_{j}\} $.  Among $_5C_2=\frac{5!}{2!3!}$ choices, we should pick up the cases where $2 \times 2$ coefficient matrix for the unknown $\{\tilde{J},\tilde{W}, \tilde{B}\}$  is regular and each of two solutions is real. 
For example, there is a case where $\tilde{J}_{3}$, and $\tilde{W}_2$ are independent real variables with others zero, 
such that  Eq.(\ref{aku2x2}) can be reduced to 
\begin{eqnarray}\label{aku2x3}
&& i\hbar \frac{\partial  C_4}{\partial R}  =   \tilde{J}_3 C_4 -i 2 \tilde{W}_2  C_2 \\\nonumber
&&
i\hbar \frac{\partial  C_2}{\partial R} = i 2 \tilde{W}_2 C_4 -\tilde{J}_3 C_2 .
\end{eqnarray}
Equation (\ref{aku2x3}) has a solution:
\begin{equation}
\tilde{J}_3= \frac{a C_4+b C_2}{C_4^2-C_2^2}=0
\end{equation}
\begin{equation}
\tilde{W}_2=\frac{i (a C_2+b C_4)}{2 \left(C_2^2-C_4^2\right)},
\end{equation}
where $a= i \hbar \frac{\partial C_4}{\partial R}$,  $b=i \hbar \frac{\partial C_2}{\partial R}$. Noting Eq.(\ref{bor}), we find $\tilde{J}_3 =0$. 

We find that each solution consists of 2 real variables with one given by $\tilde{W}_2$ and 
the other one from 4 candidates $(\tilde{J}_1,\tilde{J}_2,\tilde{J}_3,\tilde{B}_x)$ responsible 
to the real part of $\tilde{\mathcal{H}}$. Other 3 solutions of Eq.(\ref{aku2x2}) are available in a similar way, whose expressions are \\
($\tilde{B}_x$ =  0, 
$\tilde{W_2}$ = $ \frac{i (a C_4-b C_2)}{4 C_2 C_4}$),\\
($\tilde{J}_1$ =  0, 
$\tilde{W_2}$ = $\frac{i (a C_2-b C_4)}{2 \left(C_2^2+C_4^2\right)}$), \\
and\\
($\tilde{J}_2$ =  0, 
$\tilde{W_2}$ = $\frac{i (a C_2+b C_4)}{2 \left(C_2^2-C_4^2\right)}$).

Using the explicit expressions for $C_2, C_4$ of the ground state and their derivatives, however, 
the above 4 solutions turn out to be degenerate, having the identical the value  
\begin{equation}\label{final}
\tilde{W}_2 =\frac{-J \frac{\partial B_x}{\partial R}+B_x \frac{\partial J}{\partial R}}{4 \left(B_x^2+J^2\right)} 
\end{equation}
with all other interactions vanishing.
The state-dependent counter-diabatic terms and the fast forward Hamiltonian are written respectively as
\begin{equation}\label{AP-CD}
\mathcal{H} = \begin{pmatrix} 
0 &-iv(t)\tilde{W}_2 &-iv(t)\tilde{W}_2 &0  \\
 iv(t)\tilde{W}_2 & 0 & 0 & iv(t)\tilde{W}_2 \\
 iv(t)\tilde{W}_2 & 0 & 0 & iv(t)\tilde{W}_2 \\
  0 &-iv(t)\tilde{W}_2 &-iv(t)\tilde{W}_2 & 0
 \end{pmatrix} ,
\end{equation}

\begin{eqnarray}\label{HFF-ML}
H_{FF}&=& J(R(\Lambda(t))) \sigma_1^z \sigma_2^z - \frac{1}{2}(\sigma_1^x+ \sigma_2^x)B_x(R(\Lambda(t)))\nonumber\\
&+&v(t)\tilde{W}_2(R(\Lambda(t)))(\sigma_1^y \sigma_2^z +\sigma_1^z \sigma_2^y).
\end{eqnarray}

Choosing another eigenstate corresponding to the highest eigenvalue $\sqrt{J^2+B_x^2}$ below Eq. (\ref{Ap-matr}), we can reproduce the  regularization term  in Eq.(\ref{final}) and the counter-diabatic term  in Eq.(\ref{AP-CD}), and therefore these terms are state-independent. 

By applying Demirplak-Rice-Berry formula in Eq.(\ref{ric}), on the other hand, Opatrn\`{y} and  M\o{}lmer \cite{24} obtained the state-independent counter-diabatic terms $\mathcal{H}$ which agrees with Eq.(\ref{AP-CD}). In fact, using the polar coordinate $J= \rho \sin \phi$ and $B_x$= $\rho \cos \phi$, Eq.(\ref{final}) reduce to $\frac{1}{4} \frac{\partial \phi}{\partial R}$, and therefore the counter-diabatic term is described by $W_2$ = $\frac{d R}{dt} \tilde{W}_2$ = $\frac{1}{4} \dot{\phi}$ \cite{24}.\\[1cm]

\subsection{Quantum annealing model}\label{3a}
The spin analogue of the quantum annealing was proposed by Kadowaki and Nishimori \cite{20}, and has received a wide attention in the context of quantum computing \cite{21}. It should be noted: Our interest lies in showing a variety of
driving fields or counter-diabatic terms for two-spin systems, and more practical subjects, such as finding the ground state of many-spin systems described by a very complicated Hamiltonian and applying the fast-forward protocol to accelerate the quantum adiabatic computation when the final ground state is unknown, are outside of the scope of the present work.

The minimum (two-spin) Hamiltonian here is written as
\begin{equation}
H_0 = -J \sigma_1^z \sigma_2^z - \frac{1}{2}(\sigma_1^z +\sigma_2^z)B_z- \frac{1}{2}(\sigma_1^x+ \sigma_2^x)B_x,
\end{equation}
where $J$ and $B_z$ are positive constants, and $B_x=B_x(R(t))$ plays the role of tunneling among spin up and down states. By decreasing $B_x$ 
from a large positive value towards $0$, the entangled state tends to the ground state of the Ising model.
Arranging the bases as $\Ket{\uparrow\uparrow} $, $\Ket{\uparrow\downarrow} $, $\Ket{\downarrow\uparrow}$, and $\Ket{\downarrow\downarrow}$, we obtain 
\begin{equation}
H_0=\left(
\begin{array}{cccc}
 -J-B_z & -\frac{B_x}{2} & -\frac{B_x}{2} & 0 \\
 -\frac{B_x}{2} & J & 0 & -\frac{B_x}{2} \\
 -\frac{B_x}{2} & 0 & J & -\frac{B_x}{2} \\
 0 & -\frac{B_x}{2} & -\frac{B_x}{2} & -J+B_z \\
\end{array}
\right).
\end{equation}
The eigenvalues are 
\begin{eqnarray}
\lambda_1 &=& J,\nonumber\\
\lambda_2 &=&-\frac{J}{3} +\beta+\bar{\beta},\nonumber\\
\lambda_3 &=&-\frac{J}{3} -\frac{1}{2 } (\beta+\bar{\beta}) -  \frac{i \sqrt{3}}{2}(\bar{\beta}-\beta),\nonumber\\
\lambda_4 &=&-\frac{J}{3} -\frac{1}{2 } (\beta+\bar{\beta}) +  \frac{i \sqrt{3}}{2}(\bar{\beta}-\beta),
\end{eqnarray}
where
\begin{eqnarray}\label{QA-alp-beta}
\beta &=& \sqrt[3]{\sqrt{\gamma_{-}^2-\gamma_{+}^3}+\gamma_{-} }, \nonumber\\
\gamma_{+} &=&\frac{B_x^2}{3}+\frac{B_z^2}{3}+\frac{4 J^2}{9}, \nonumber\\
\gamma_{-} &=& \frac{B_x^2 J}{3}-\frac{2 B_z^2 J}{3}+\frac{8 J^3}{27}.
\end{eqnarray}
All eigenvalues above are real. 
The eigenvector for the ground state (with the eigenvalue $\lambda_3$) is 
\begin{eqnarray}\label{C11}
C_1&=&\zeta  \left(-\frac{B_x^2-2 B_zJ+2 J^2}{B_x^2}+\frac{2 B_z \Gamma }{B_x^2}+\frac{2\Gamma ^2}{B_x^2}\right), \nonumber\\
C_2&=&\zeta  \left(\frac{B_z-J}{B_x}+\frac{\Gamma }{B_x}\right), \nonumber\\
C_3&=&\zeta  \left(\frac{B_z-J}{B_x}+\frac{\Gamma }{B_x}\right), \nonumber\\
C_4 &=& \zeta,
\end{eqnarray}
where
\begin{equation}
\Gamma = \frac{1}{2}i \sqrt{3} (\bar{\beta}-\beta)+\frac{1}{2 } (\beta+\bar{\beta})+\frac{J}{3}, 
\end{equation}
and
$\zeta$ is normalization factor written as
\begin{equation}
\zeta =\frac{1}{\sqrt{\left(-\frac{B_x^2-2 B_zJ+2 J^2}{B_x^2}+\frac{2 B_z \Gamma }{B_x^2}+\frac{2\Gamma ^2}{B_x^2}\right)^2+2 \left(\frac{B_z-J}{B_x}+\frac{\Gamma }{B_x}\right)^2+1}}.
\end{equation}

We shall concentrate on the fast forward of the quasi-adiabatic dynamics of the ground state with the eigenvalue $\lambda_3$. From the eigenvector we see,  $C_2=C_3$, 
and $C_1$, $C_2$, $C_3$, and $C_4$ are real. From the normalization ($C_1^2+2 C_2^2+ C_4^2 =1$), we see
\begin{equation}\label{berry}
C_1 \frac{\partial C_1}{\partial R}+ 2 C_2 \frac{\partial C_2}{\partial R}+C_4 \frac{\partial C_4}{\partial R} =0,
\end{equation}
and then adiabatic phase ($\xi$) is equal to 0. Because of the symmetry ($C_3 = C_2$), the equation for the regularization terms ($\mathcal{\tilde{H}}$) in Eq.(\ref{sum2}) is written as
\begin{eqnarray}\label{aku3}
&& i\hbar \frac{\partial  C_1}{\partial R}  =   \tilde{\mathcal{H}}_{11}C_1 + (\tilde{\mathcal{H}}_{12} +\tilde{\mathcal{H}}_{13}) C_2+\tilde{\mathcal{H}}_{14}C_4, \nonumber\\
&&
i\hbar \frac{\partial  C_2}{\partial R} =                \tilde{\mathcal{H}}_{21}C_1 +  (\tilde{\mathcal{H}}_{22}+ \tilde{\mathcal{H}}_{23})C_2 +\tilde{\mathcal{H}}_{24} C_4,  \nonumber \\
&& i\hbar \frac{\partial  C_2}{\partial R} =                \tilde{\mathcal{H}}_{31}C_1 +  (\tilde{\mathcal{H}}_{32}+ \tilde{\mathcal{H}}_{33})C_2 +\tilde{\mathcal{H}}_{34} C_4,   \nonumber \\ 
&&
 i\hbar \frac{\partial  C_4}{\partial R}  =    \tilde{\mathcal{H}_{41}}C_1+ (\tilde{\mathcal{H}}_{42} + \tilde{\mathcal{H}}_{43})C_2 +  \tilde{\mathcal{H}}_{44}C_4. \nonumber \\
\end{eqnarray}
Noting that $\mathcal{\tilde{H}}_{21}$ = $\mathcal{\tilde{H}}_{31}$, $\mathcal{\tilde{H}}_{24}$ = $\mathcal{\tilde{H}}_{34}$ and $\mathcal{\tilde{H}}_{22}$ + $\mathcal{\tilde{H}}_{23}$ = $\mathcal{\tilde{H}}_{32}$ +$\mathcal{\tilde{H}}_{33}$, we find that the 2nd and 3rd lines are degenerate. Then the independent equations in Eq.(\ref{aku3}) reduce to
\begin{eqnarray}\label{aku4}
&& i\hbar \frac{\partial  C_1}{\partial R}  =   \tilde{\mathcal{H}}_{11}C_1 + \tilde{\mathcal{A}}_{1}  C_2+\tilde{\mathcal{H}}_{14}C_4, \nonumber \\
&&
i\hbar \frac{\partial  C_2}{\partial R} =                \tilde{\mathcal{H}}_{21} C_1 +  \tilde{\mathcal{A}}_{2}C_2+\tilde{\mathcal{H}}_{24}C_4, \nonumber  \\
&&
i\hbar \frac{\partial  C_4}{\partial R} =                \tilde{\mathcal{H}}_{41} C_1 +  \tilde{\mathcal{A}}_{4}C_2+\tilde{\mathcal{H}}_{44}C_4,
\end{eqnarray}
where $\tilde{\mathcal A}_1 \equiv \tilde{\mathcal H}_{12} +  \tilde{\mathcal H}_{13}$, $\tilde{\mathcal A}_2 \equiv \tilde{\mathcal H}_{22} +  \tilde{\mathcal H}_{23}$, and $\tilde{\mathcal A}_4 \equiv \tilde{\mathcal H}_{42} +  \tilde{\mathcal H}_{43}$. 

To solve the ternary simultaneous linear equations for  $\{\tilde{\mathcal H}_{ij}\}$  in Eq.(\ref{aku4}), we should note the nature of $\tilde{\mathcal H}$ in the candidate for regularization terms in Eq.(\ref{cdterm}), which is Hermitian and traceless and has other symmetries. 
This setup implies that we can choose three independent real variables out of nine real variables in Eq.(\ref{cdterm}). Among $_9C_3 = \frac{9 !}{3 ! 6 !}$ choices, however, we should pick up only  the cases where 3 $\times$ 3 coefficient matrix for the unknown $\{\tilde{J}, \tilde{W},\tilde{B}\}$ is regular and each of three solutions is real. For example, there is a choice where $\tilde{B}_z$, $\tilde{B}_y$, and $\tilde{W}_2$ are independent real variables with others zero, such that Eq.(\ref{aku4}) can simply be rewritten as
\begin{eqnarray}\label{aku10}
&& i\hbar \frac{\partial  C_1}{\partial R}  =   \tilde{B}_z C_1 -i (\tilde{B}_y+2\tilde{W}_2) C_2 ,   \nonumber \\
&&
i\hbar \frac{\partial  C_2}{\partial R} = i(\frac{\tilde{B}_y}{2}+\tilde{W}_2) C_1 +  i(\tilde{W}_2-\frac{\tilde{B}_y}{2})C_4 , \nonumber\\
&&
i\hbar \frac{\partial  C_4}{\partial R} = i(\tilde{B}_y-2 \tilde{W}_2)C_2-\tilde{B}_z C_4.
\end{eqnarray}
Then solving Eq.(\ref{aku10}), we obtain
\begin{eqnarray}\label{ByW2}
\tilde{B}_z &=& \frac{a C_1+2 b C_2+c C_4}{(C_1-C_4) (C_1+C_4)} = 0, \nonumber\\
\tilde{B}_y &=& -\frac{i (a C_4+2 b C_2+c C_1)}{2 C_2 (C_1-C_4)}, \nonumber\\
\tilde{W}_2 &=& -\frac{i (-a C_4+2 b C_2-c C_1)}{4 C_2 (C_1+C_4)}, 
\end{eqnarray}
where $a = i \hbar\frac{\partial C_1}{\partial R}$, $b =i \hbar \frac{\partial C_2}{\partial R}$, and $c = i \hbar \frac{\partial C_4}{\partial R}$. $\tilde{W}_2$ is responsible to $ (\sigma_1^y\sigma_2^z+\sigma_1^z\sigma_2^y)$. Noting Eq.(\ref{berry}) we find $\tilde{B}_z$ =0. 
The regularization terms and the fast forward Hamiltonian are respectively written as 
\begin{equation}
\mathcal{\tilde{H}} = \begin{pmatrix} 
0 &-i\frac{\tilde{B}_y}{2}-i\tilde{W}_2 &-i\frac{\tilde{B}_y}{2}-i\tilde{W}_2 & 0  \\
 i\frac{\tilde{B}_y}{2}+i\tilde{W}_2 & 0 & 0 & -i\frac{\tilde{B}_y}{2}+i\tilde{W}_2 \\
 i\frac{\tilde{B}_y}{2}+i\tilde{W}_2 & 0 & 0 & -i\frac{\tilde{B}_y}{2}+i\tilde{W}_2 \\
  0 &i\frac{\tilde{B}_y}{2}-i\tilde{W}_2 &i\frac{\tilde{B}_y}{2}-i\tilde{W}_2 & 0
 \end{pmatrix}
\end{equation}
and
\begin{eqnarray}\label{HFF}
H_{FF}&=&-J \sigma_1^z \sigma_2^z - \frac{1}{2}(\sigma_1^z +\sigma_2^z)B_z \nonumber\\
&-& \frac{1}{2}(\sigma_1^x+ \sigma_2^x)B_x(R(\Lambda(t))) \nonumber\\
&+&v(t)\tilde{W}_2(R(\Lambda(t)))( \sigma_1^y \sigma_2^z +  \sigma_1^z \sigma_2^y) \nonumber\\
&+&\frac{1}{2}(\sigma_1^y +\sigma_2^y)v(t)\tilde{B}_y(R(\Lambda(t))) 
\end{eqnarray}
where $\hbar$ =1. 

We find that each solution consists of 3 real variables, among which 2 come from 3 candidates ($\tilde{B}_y, \tilde{W}_1, \tilde{W}_2$) responsible to the imaginary part of $\mathcal{\tilde{H}}$ 
in Eq.(\ref{cdterm}) and 1 comes from 6 candidates responsible to its real part. Therefore the total number of solutions is $_3C_2 \times  _6C_1=18$. Other 17 solutions of Eq.(\ref{aku4}) are also available as above. 
All of 18 solutions are listed in Table \ref{table1} in Appendix B, which, multiplied by $v(t)$, are counter-diabatic terms proper to the ground-state eigenvector  in Eqs.(\ref{C11}).

\begin{figure}[!h]
\subfloat[]{%
  \includegraphics[width=2.5in]{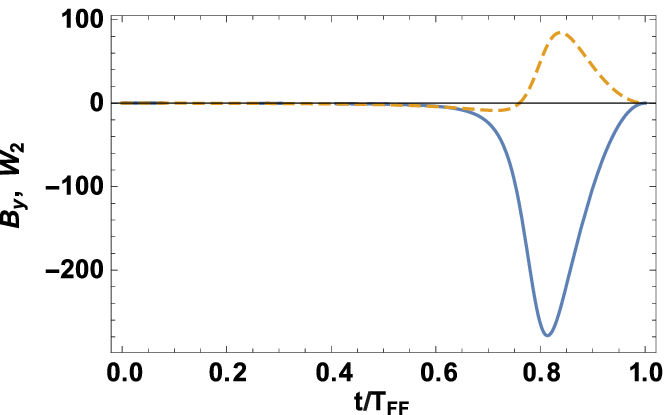}%
}
\hfill
\subfloat[]{%
  \includegraphics[width=2.5in]{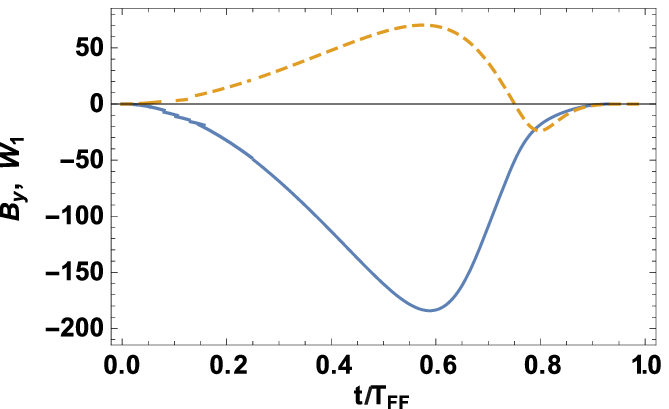}%
}
\hfill
\subfloat[]{%
  \includegraphics[width=2.5in]{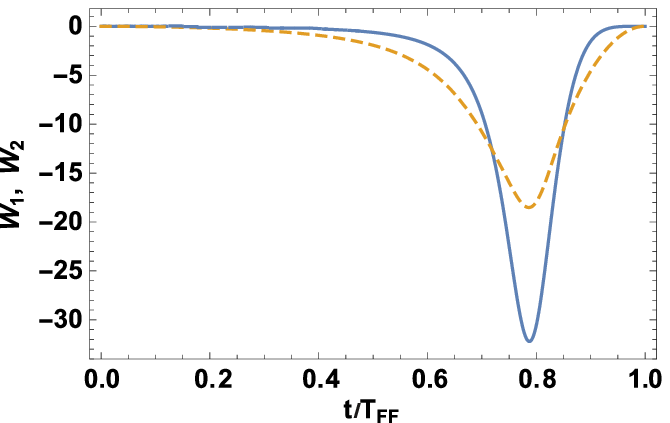}%
}
\caption{Time dependence of 3 solutions for the state-dependent counter-diabatic interactions: (a) $B_y(t)=v(t) \tilde{B}_y(R(\Lambda(t)))$ (solid line), $W_2(t)=v(t) \tilde{W}_2(R(\Lambda(t)))$ (dashed line); (b) $B_y(t)=v(t) \tilde{B}_y(R(\Lambda(t)))$ (solid line), $W_1(t)=v(t) \tilde{W}_1(R(\Lambda(t)))$ (dashed line); (c) $W_1(t)=v(t) \tilde{W}_1(R(\Lambda(t)))$ (solid line), $W_2(t)=v(t) \tilde{W}_2(R(\Lambda(t)))$ (dashed line).} \label{fig1}
\end{figure}

Using the explicit expressions for $C_1, C_2, C_4$ in Eq.(\ref{C11}) and their derivatives, however, 18 solutions have proved to be classified into 3 groups: In the first group, $(\tilde{B}_y, \tilde{W}_2)$ has the same nonzero values with others zero. Similarly, in the second  and third groups, $(\tilde{B}_y, \tilde{W}_1)$ and $(\tilde{W}_1, \tilde{W}_2)$ play such a role, respectively. To conclude,
we have 3 independent counter-diabatic terms, whose time dependence such as 
$(W_2, B_y)= (v(t)\tilde{W}_2(R(\Lambda(t))), v(t) \tilde{B}_y(R(\Lambda(t))))$ is shown in Fig. \ref{fig1} .

\begin{figure}[!h]
\subfloat[]{%
  \includegraphics[width=2.5in]{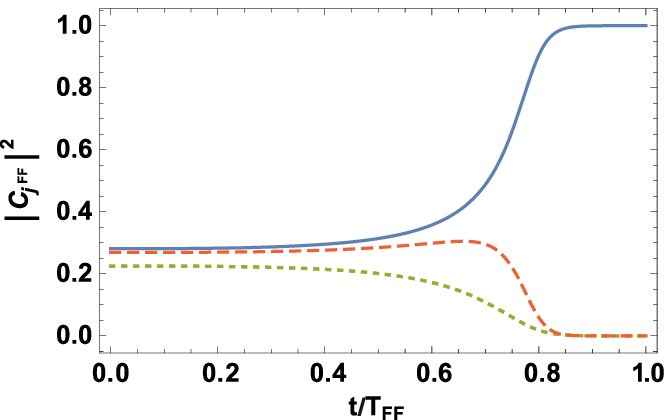}%
}
\hfill
\subfloat[]{%
  \includegraphics[width=2.5in]{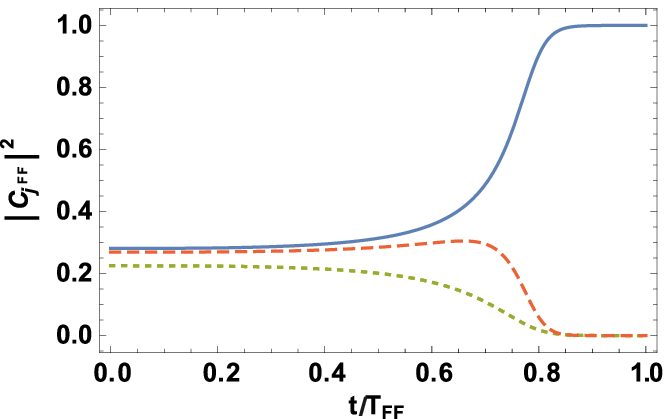}%
}
\caption{The time dependence of $|C_1^{FF}|^2$ (solid line), $|C_2^{FF}|^2$ (dotted line), $|C_3^{FF}|^2$ (dotted line), and $|C_4^{FF}|^2$ (dashed line):(a)  Obtained by solving TDSE; (b) Obtained from the eigenvector.} \label{fig2}
\end{figure}

In the fast forward Hamiltonian $H_{FF}$ in Eq.(\ref{HFF}), the time dependence of the counter-diabatic term is explicitly shown in Fig.\ref{fig1}(a). Then
we numerically solve TDSE in Eq.(\ref{TDSE-FF}) in the case of $v(t)$ in Eq.(\ref{velo}) and $R(\Lambda(t))$ in Eq.(\ref{lambda2}). Here we put $J=1$, $B_z = 0.1$,  $B_x=B_0 - R(\Lambda(t))$, $B_0 = 10 (R_0=0)$, $\bar{v}$ = 100, $T_{FF} = 0.1$. The initial state is a linear combination of $\Ket{\uparrow\downarrow} $, $\Ket{\downarrow\uparrow}$, and  $\Ket{\downarrow\downarrow} $ states, and as $B_x$ is decreased the system falls into the nonentangled (product) state $\Ket{\uparrow\uparrow}$.  
Figure \ref{fig2} shows that the initial entangled state ($C_1 = 0.5300$, $C_2 = 0.4744$, $C_3 =0.4744$, $C_4$ =0.5184 ) rapidly changes to the product state $\Ket{\uparrow\uparrow} $, i.e, the ground state of the Ising model. Figure \ref{fig2}(a) is the result of TDSE and exactly agrees with the time dependence of the eigenstate in Eqs.(\ref{C11}), depicted in Fig.\ref{fig2}(b). 
The time-dependent fidelity of the wavefunction solution $\Psi_{FF}(t)$ of TDSE in Eq.(\ref{TDSE-FF}) to the eigenfunction $\Psi_0(R(\Lambda(t)))$ in Eq.(\ref{psi0}) is defined by 
$|\Psi_{FF}^{\dag}(t) \cdot \Psi_0(R(\Lambda(t)))|=|\sum_{j=1}^4C^*_{FF,j}(t)C_j(R(\Lambda(t)))|$ in case of $N=4$. Concerning Figs.\ref{fig2}(a) and (b), we numerically confirmed the  fidelity$=1-\epsilon$ 
with $ 0 \leq \epsilon \leq 10^{-6}$ during the fast-forward time range $0 \leq t \leq T_{FF}$.

In the case of other 2 solutions whose counter-diabatic interactions are shown in Figs.\ref{fig1} (b) and (c), we also investigated TDSE numerically, and confirmed the same high fidelity of wavefunctions as in the case of Fig.\ref{fig2}. 

We further solved Eq.(\ref{sum2}) with use of the guiding Hamiltonian in Eq.(\ref{cand-reg}) in the cases of other 3 eigenvectors. However, we could not find a solution common to all four eigenvectors, namely, we could not see the state-independent counter-diabatic term. Although we investigated a more general guiding Hamiltonian including antisymmetric interactions like $\sigma_i^x\sigma_j^y-\sigma_i^y\sigma_j^x$, we  found no new result: the corresponding regularization interaction proved to be vanishing. This fact suggests the state-independent counter-diabatic term would be beyond the scope of the guiding Hamiltonian in Eq.(\ref{cand-reg}) acceptable in magnetic systems.

\subsection{Model for generation of entangled state}\label{3b}
There is a prominent model to generate the entangled state, which is the Ising model with general magnetic field \cite{22,23}. The corresponding Hamiltonian is given by:
\begin{equation}\label{GENT}
H_0 = J \sigma_1^z \sigma_2^z + \frac{1}{2}(\bm{\sigma}_1+\bm {\sigma}_2) \cdot \mathbf{B},
\end{equation}
which can generate an entangled state from the product state. In Eq.(\ref{GENT}) $\mathbf{B}=(B_x, B_y, B_z)$ with $B_z=B_z(R(t))$. $B_x, B_y$ and $J$ are assumed constants. Arranging the bases as $\Ket{\uparrow\uparrow} $, $\Ket{\uparrow\downarrow} $, $\Ket{\downarrow\uparrow}$, and $\Ket{\downarrow\downarrow}$, we obtain 
\begin{equation}
H_0=\left(
\begin{array}{cccc}
 J+B_z & \frac{B_x}{2}-i \frac{B_y}{2} & \frac{B_x}{2}-i \frac{B_y}{2} & 0 \\
\frac{B_x}{2}+i \frac{B_y}{2} & -J & 0 &\frac{B_x}{2}-i \frac{B_y}{2} \\
\frac{B_x}{2}+i \frac{B_y}{2} & 0 & -J & \frac{B_x}{2}-i \frac{B_y}{2} \\
 0 & \frac{B_x}{2}+i \frac{B_y}{2} & \frac{B_x}{2}+i \frac{B_y}{2} & J-B_z \\
\end{array}
\right).
\end{equation}
The eigenvalues are 
\begin{eqnarray}\label{QEN-energy}
\lambda_1 &=& J,\nonumber\\
\lambda_2 &=& \beta + \bar{\beta}  +\frac{J}{3},\nonumber\\
\lambda_3 &=& -\frac{1}{2} \sqrt{3}i(\beta-\bar{\beta})-\frac{1 }{2}(\beta + \bar{\beta} )+\frac{J}{3},\nonumber\\
\lambda_4 &=& \frac{1}{2} \sqrt{3} i (\beta-\bar{\beta})-\frac{1 }{2}(\beta + \bar{\beta} )+\frac{J}{3},
\end{eqnarray}
where
\begin{eqnarray}\label{EN-alp-beta}
\beta &=& \sqrt[3]{\sqrt{\gamma_{-}^2-\gamma_{+}^3}+\gamma_{-} } \quad, \nonumber\\
\gamma_{+} &=&\frac{B_z^2}{3}+\frac{4 |Z|^2}{3}+\frac{4 J^2}{9} , \nonumber\\
\gamma_{-} &=& \frac{2 B_z^2 J}{3}-\frac{4 J |Z|^2}{3}-\frac{8 J^3}{27}.
\end{eqnarray}
All eigenvalues above are real. The eigenvector for the ground state (with the eigenvalue $\lambda_4$) is 
\begin{eqnarray}\label{C1}
C_1&=& \frac{\zeta  \left(B_z \Gamma -\left(-B_z J+2 |Z|^2+J^2\right)+\Gamma ^2\right)}{2 \left(Z^*\right)^2}, \nonumber\\
C_2 &=& \frac{\zeta  (B_z+\Gamma -J)}{2 Z^*}, \nonumber\\
C_3 &=& \frac{\zeta  (B_z+\Gamma -J)}{2 Z^*}, \nonumber\\
C_4 &=& \zeta,
\end{eqnarray}
with $Z=\frac{1}{2} (B_x-i B_y)$, and
\begin{equation}
\Gamma = \frac{1}{2} \sqrt{3} i (\beta-\bar{\beta})-\frac{1 }{2}(\beta + \bar{\beta} )+\frac{J}{3}.
\end{equation}
$\zeta$ is normalization factor given by 
\begin{equation}
\zeta =\frac{1}{\sqrt{\left(\frac{B_z \Gamma -(-B_z J+2 |Z|^2+J^2)+\Gamma ^2}{2 |Z|^2}\right)^2+2 \left(\frac{B_z+\Gamma -J}{2 |Z|}\right)^2+1}}.
\end{equation}
We shall concentrate on the fast forward of the quasi-adiabatic dynamics of the ground state with the eigenvalue $\lambda_4$. From the eigenvector we see,  $C_2=C_3$, and $C_1$, $C_2$, $C_3$, $C_4$ are complex. Again noting the fact that $\mathcal{\tilde{H}}_{21}$ = $\mathcal{\tilde{H}}_{31}$, $\mathcal{\tilde{H}}_{24}$ = $\mathcal{\tilde{H}}_{34}$ and $\mathcal{\tilde{H}}_{22}$ + $\mathcal{\tilde{H}}_{23}$ = $\mathcal{\tilde{H}}_{32}$ +$\mathcal{\tilde{H}}_{33}$, Eq.(\ref{sum2}) becomes three independent equations :
\begin{eqnarray}\label{aku5}
&& i\hbar (\frac{\partial  C_1}{\partial R}- L C_1)  =   \tilde{\mathcal{H}}_{11}C_1 + \tilde{\mathcal{A}}_{1}  C_2+\tilde{\mathcal{H}}_{14}C_4 ,\nonumber\\
&&
i\hbar (\frac{\partial  C_2}{\partial R}- L C_2) =                \tilde{\mathcal{H}}_{21} C_1 +  \tilde{\mathcal{A}}_{2}C_2+\tilde{\mathcal{H}}_{24}C_4 ,\nonumber\\
&&
i\hbar (\frac{\partial  C_4}{\partial R}-L C_4) =                \tilde{\mathcal{H}}_{41} C_1 +  \tilde{\mathcal{A}}_{4}C_2+\tilde{\mathcal{H}}_{44}C_4, \nonumber\\
\end{eqnarray}
where $\tilde{A}_1$ = $\mathcal{\tilde{H}}_{12}$ + $\mathcal{\tilde{H}}_{13}$, $\tilde{A}_2$ = $\mathcal{\tilde{H}}_{22}$ + $\mathcal{\tilde{H}}_{23}$, $\tilde{A}_4$ = $\mathcal{\tilde{H}}_{42}$ + $\mathcal{\tilde{H}}_{43}$, and
\begin{equation}
L= C_1^* \frac{\partial C_1}{\partial R}+ 2  C_2^* \frac{\partial C_2}{\partial R} +  C_4^* \frac{\partial C_4}{\partial R}.
\end{equation}
We shall take a similar procedure as in  the previous sub-Sections:
To solve the ternary simultaneous linear equations for  $\{\tilde{\mathcal H}_{ij}\}$  in Eq. (\ref{aku5}), we should choose three independent real variables out of nine real variables in Eq.(\ref{cdterm}). Then only the cases should be picked up where 3$\times$3 coefficient matrix for the unknown $\{\tilde{J}, \tilde{W},\tilde{B}\}$  is regular and each of three solutions are real.
There exists a choice where  $\tilde{W}_3$, $\tilde{B}_y$, and $\tilde{W}_1$ are independent real variables with others zero. 
Then Eq.(\ref{aku5}) can be cast into the form
\begin{eqnarray}\label{aku8}
&& i\hbar \frac{\partial  C_1}{\partial R}-L C_1  =  (-i\tilde{B}_y+2\tilde{W}_3) C_2-2 i \tilde{W}_1 C_4,   \nonumber \\
&&
i\hbar \frac{\partial  C_2}{\partial R}- L C_2 = (i\tilde{B}_y+\tilde{W}_3) C_1 +  (-i\tilde{B}_y-\tilde{W}_3)C_4, \nonumber  \\
&&
i\hbar \frac{\partial  C_4}{\partial R} - L C_4 = 2 i\tilde{W}_1 C_1+ (i\tilde{B}_y- 2\tilde{W}_3)C_2.
\end{eqnarray}
Solving Eq.(\ref{aku8}), we obtain
\begin{eqnarray}\label{CD21}
\tilde{W}_3 &=& \frac{a C_1+2 b C_2+c C_4}{4 C_2 \left(C_1-C_4\right)}, \nonumber \\
\tilde{B}_y &=& -\frac{i \left(-a C_1+2 b C_2-c C_4\right)}{2 C_2 \left(C_1-C_4\right)}, \nonumber \\
\tilde{W}_1 &=& -\frac{i (a+c)}{2 \left(C_1-C_4\right)},
\end{eqnarray}
where 
 $a = i(\frac{\partial C_1}{\partial R}- L C_1)$, $b =i (\frac{\partial C_2}{\partial R}- L C_2)$, and $c = i(\frac{\partial C_4}{\partial R}- L C_4) $.  The regularization term and the fast forward Hamiltonian are respectively written as 
\begin{widetext}
\begin{equation}
\mathcal{\tilde{H}} = \begin{pmatrix} 
0 &-i\tilde{B}_y+\tilde{W}_3 &-i\tilde{B}_y+ \tilde{W}_3 & -2 i \tilde{W}_1  \\
i\tilde{B}_y+\tilde{W}_3 & 0 & 0 & -i\tilde{B}_y-\tilde{W}_3 \\
 i\tilde{B}_y+\tilde{W}_3 & 0 & 0 & -i\tilde{B}_y-\tilde{W}_3 \\
  2 i  \tilde{W}_1 & i\tilde{B}_y-\tilde{W}_3 & i\tilde{B}_y-\tilde{W}_3 & 0
 \end{pmatrix},
\end{equation}
\begin{eqnarray}\label{HFF-EN}
H_{FF}&=&J \sigma_1^z \sigma_2^z + \frac{1}{2}(\bm{\sigma}_1+\bm {\sigma}_2) \cdot \mathbf{B} \nonumber\\
&+&v(t)\tilde{W}_1(R(\Lambda(t)))(\sigma_1^x \sigma_2^y +\sigma_1^y \sigma_2^x)+v(t)\tilde{W}_3(R(\Lambda(t)))(\sigma_1^z \sigma_2^x+ \sigma_1^x \sigma_2^z) + \frac{1}{2}(\sigma_1^y +\sigma_2^y)v(t)\tilde{B_y}(R(\Lambda(t))),\nonumber\\
\end{eqnarray}
where $\mathbf{B}=(B_x, B_y, B_z(R(\Lambda(t))))$.
\end{widetext}

In the model for generation of entangled states, we find only two solutions available. Another solution for the regularization term is:
\begin{eqnarray}\label{CD24}
\tilde{B}_x &=& \frac{a C_1+2 b C_2+c C_4}{2 C_2 \left(C_1+C_4\right)}, \nonumber\\
\tilde{W}_2 &=& -\frac{i \left(-a C_1+2 b C_2-c C_4\right)}{4 C_2 \left(C_1+C_4\right)}, \nonumber\\
\tilde{W}_1 &=& \frac{i (a-c)}{2 \left(C_1+C_4\right)}.
\end{eqnarray}

\begin{figure}[!h]
\subfloat[]{%
  \includegraphics[width=2.5in]{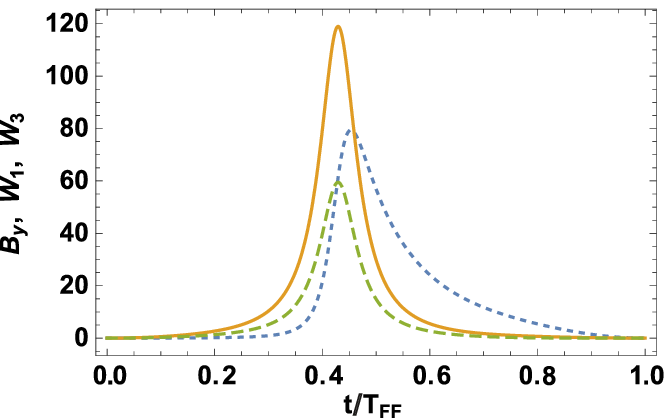}%
}
\hfill
\subfloat[]{%
  \includegraphics[width=2.5in]{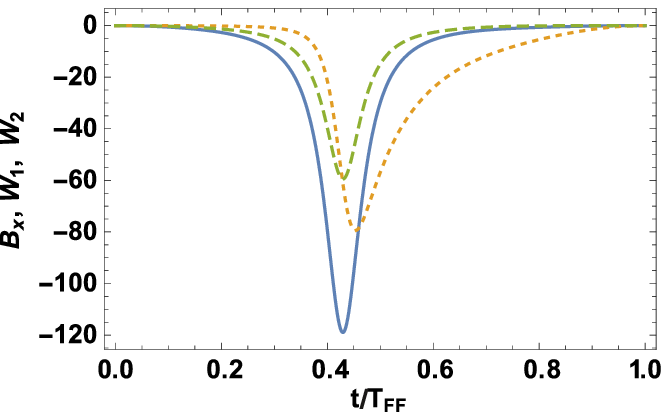}%
}
\caption{Time dependence of 2 solutions for the state-dependent counter-diabatic interactions: 
(a) $B_y(t)=v(t) \tilde{B}_y(R(\Lambda(t)))$ (solid line), $W_1(t)=v(t) \tilde{W}_1(R(\Lambda(t)))$ (dotted line), 
and $W_3(t)=v(t) \tilde{W}_3(R(\Lambda(t)))$ (dashed line); (b) $B_x(t)=v(t) \tilde{B}_x(R(\Lambda(t)))$ (solid line), $W_1(t)=v(t) \tilde{W}_1(R(\Lambda(t)))$ (dotted line), and $W_2(t)=v(t) \tilde{W}_2(R(\Lambda(t)))$ (dashed line).} \label{fig3}
\end{figure}

The dynamics of $(\tilde{W}_3, \tilde{B}_y,\tilde{W}_1)$  in Eq.(\ref{CD21}) and  $(\tilde{B}_x, \tilde{W}_2, \tilde{W}_1)$  in Eq.(\ref{CD24}) multiplied by $v(t)$ are shown in Fig.\ref{fig3} (a) and (b), respectively.

\begin{figure}[!h]
\subfloat[]{%
  \includegraphics[width=2.5in]{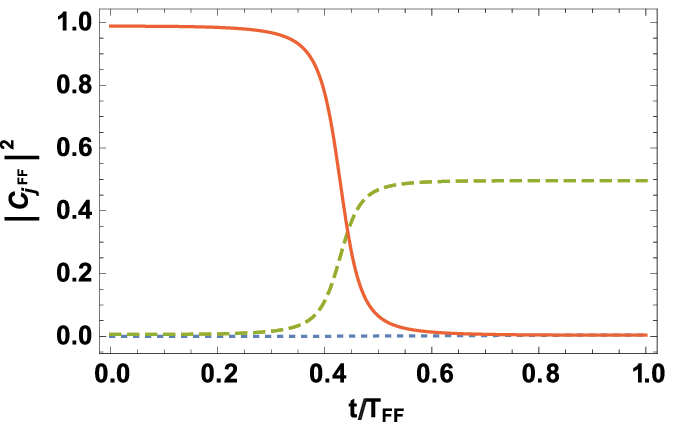}%
}
\hfill
\subfloat[]{%
  \includegraphics[width=2.5in]{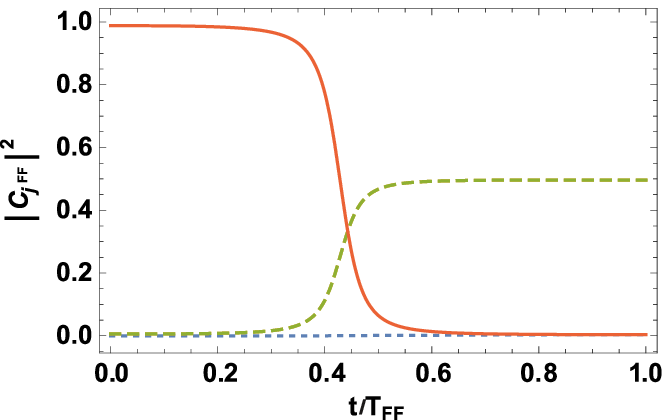}%
}
\caption{The time dependence of $|C_4^{FF}|^2$ (solid line), $|C_2^{FF}|^2$ (dashed line), $|C_3^{FF}|^2$ (dashed line), and $|C_1^{FF}|^2$ (dotted line):(a)  Obtained by solving TDSE ; (b)  Obtained from eigenvectors.} \label{fig4}
\end{figure}

In the case of the solution in Eq.(\ref{CD21}) whose behavior is shown in Fig.\ref{fig3}(a),
we numerically solve TDSE in Eq. (\ref{TDSE-FF}) with $H_{FF}$ in Eq. (\ref{HFF-EN}) in the case of $v(t)$ in Eq.(\ref{velo}) and $R(\Lambda(t))$ in Eq.(\ref{lambda2}).
Putting $J=4(B_x^2+B_y^2)$, $B_x=1$, $B_y=1$, $B_z = B_0-R(\Lambda(t))$,  $B_0 = 25(R_0=0)$, $\bar{v}$ = 250, $T_{FF} = 0.1$,  we see in Fig.\ref{fig4}(a) the dynamics shows a change from a nonentangled state at $t= 0$ where only $C_4$ appears to the entangled state at $t=T_{FF}$ where only $C_2$ and $C_3$  appear. In fact the initial product state $(C_4= 1, C_2=C_3=C_1=0)$ rapidly changes to the entangled state $(C_1 = C_4 = 0,  C_2=C_3=\frac{1}{\sqrt{2}})$. 

Figure \ref{fig4}(a) is exactly the same as the temporal change of the ground state defined in Eq.(\ref{C1}) 
shown in Fig.\ref{fig4}(b).
We numerically evaluated the time-dependent fidelity of the wavefunction solution $\Psi_{FF}$ to the eigenfunction $\Psi_0$, and found the fidelity$=1-\epsilon$ with 0  $\leq \epsilon \leq 10^{-6}$.

In the case of another solution in Eq.(\ref{CD24}), we also investigated TDSE numerically, and confirmed the same high fidelity of wavefunction as in the case of Fig.\ref{fig3}. 
Two solutions in Eqs.(\ref{CD21}) and (\ref{CD24}), multiplied by $v(t)$, are counter-diabatic terms proper 
to the particular eigenvector in Eq.(\ref{QEN-energy}).

We further solved Eq. (\ref{sum2}) with use of the guiding Hamiltonian in Eq.(\ref{cand-reg}) in the cases of other three eigenvectors. 
However, we could not see the state-independent counter-diabatic term. Therefore, in the model for generation of entangled states, the state-independent counter-diabatic term 
would be available from a guiding Hamiltonian more general than in Eq.(\ref{cand-reg}) as mentioned in the beginning of Section \ref{geneSec} and at the end of  Section \ref{3a}.

In closing this Section, we should note about the extension of the present scheme to many-spin systems. The solution $\mathcal{\tilde{H}}_n$ available from Eq.(\ref{sum2})  is exact, in contrast to the truncated variant of $\mathcal{H}$ to be obtained from Demirplak-Rice-Berry's formula  in Eq.(\ref{ric}) by extracting only a ground-state contribution in the treatment of many-spin systems\cite{13,camp}. If there is a knowledge of the ground state of a many-spin system, we can solve Eq.(\ref{sum2}) under the strategy to use a candidate regularization Hamiltonian $\mathcal{\tilde{H}}_{n=0}$ like Eq.(\ref{cand-reg}) and expect a  variety of ground-state counter-diabatic terms.\\

\section{Conclusion}\label{concl}
By extending the idea of fast forward for adiabatic orbital dynamics, we presented a scheme of the fast forward of adiabatic spin dynamics of  quantum entangled states. We settled the quasi-adiabatic dynamics by adding the regularization terms to the original Hamiltonian and then accelerated it with use of a large time-scaling factor. Assuming the experimentally-realizable candidate Hamiltonian consisting of the exchange interactions and magnetic field, we solved the regularization terms. 
We took a strategy: a product of the mean value $\bar{\alpha}$ of an infinitely-large time-scaling factor $\alpha(t)$ and an infinitesimally-small growth rate $\epsilon$ in the quasi-adiabatic parameter should satisfy the constraint $\bar{\alpha} \cdot \epsilon$ = $finite$ in the asymptotic limit $\bar{\alpha} \rightarrow \infty$ and $\epsilon \rightarrow 0$. 
The regularization terms multiplied by the velocity function give rise to the state-dependent counter-diabatic terms. As an illustration we chose 3 systems of coupled spins whose ground states have entangled states.
Our scheme has generated a variety of fast-forward Hamiltonians characterized by state-dependent counter-diabatic terms for each of adiabatic states, which can include the state-independent one.  Broad range of choosing the driving  pair interactions and magnetic field will make flexible the experimental design of accelerating the adiabatic  quantum spin dynamics or quantum computation.

\begin{acknowledgements}
We are grateful to K. Takahashi for discussions in the early stage of this work. The work is funded by Hibah Disertasi Doktor Ristekdikti 2016. I.S. is grateful to D. Matrasulov for his great hospitality at Turin Polytechnic University in Tashkent. The work of B.E.G. is supported by PUPT Ristekdikti-ITB 2017.
\end{acknowledgements}

\renewcommand{\theequation}{A\arabic{equation}}
\section*{Appendix A.  Derivation of Eq.(\ref{TDSE-FF})}
Taking the time derivative of $\Psi_{FF}(t)$ in Eq.(\ref{psiff}) and using the equalities $\partial_t \textbf{C}(R(\Lambda(t)))=\alpha \epsilon \partial_R \textbf{C}$ and $\partial_t \xi(R(\Lambda(t)))=i\textbf{C}^{\dagger}\partial_t \textbf{C}=i\alpha\epsilon\textbf{C}^{\dagger}\partial_R \textbf{C}$, we have
\begin{eqnarray}
i\hbar\dot{\Psi}_{FF}&=&\left[ i\hbar\alpha\epsilon(\partial_R \textbf{C} -(\textbf{C}^{\dagger}\partial_R \textbf{C})\textbf{C})
+E \textbf{C}\right] \nonumber\\
&\times & e^{-\frac{i}{\hbar}\int_{0}^{t}E((R(\Lambda(t'))))dt'} e^{i\xi((R(\Lambda(t))))}.
\end{eqnarray}
The first and second terms in the angular bracket on the right-hand side are replaced by $\alpha\epsilon \mathcal{\tilde{H}}_n \textbf{C}(R(\Lambda(t)))$ and $H_0\textbf{C}(R(\Lambda(t)))$, respectively, by using Eqs.(\ref{sum2}) and (\ref{satu}). Then, using the definition of $\Psi_{FF}(t)$ and taking the asymptotic limit, we obtain Eq.(\ref{TDSE-FF}).

\begin{widetext}
\section*{Appendix B. Regularization terms of the model in Section \ref{3a}}
\begin{table}[H]
\caption{A list of formal solutions of regularization terms for the model in Section \ref{3a}. 
$C_1, C_2(=C_3)$ and $C_4$ are defined in Eq.(\ref{C11}). 
$a = i \hbar\frac{\partial C_1}{\partial R}$, $b =i \hbar \frac{\partial C_2}{\partial R}$, and $c = i \hbar \frac{\partial C_4}{\partial R}$. In each frame, variables other than listed ones are vanishing. Multiplying each term by $v(t)$ gives the driving interaction that corresponds to the state-dependent counter-diabatic term. 
For instance, $B_y$ = $v(t) \tilde{B}_y(R(\Lambda(t)))$. 18 solutions are classified to 3 groups. The 1st-6th solutions are degenerate and belongs to the 1st group. Similarly the 7th-12th ones to the 2nd group and the 13th-18th ones to the 3rd group. See details in the second paragraph below Eq.(\ref{HFF}).}

\label{table1}
\begin{center}
\begin{tabular}{|l|l|l|l|}
\hline
 & $\tilde{B_z}$ =  0 &  & $\tilde{J_2}$ =0 \\
1. &$\tilde{B_y}$ =   $-\frac{i (a C_4+2 b C_2+c C_1)}{2 C_2 (C_1-C_4)}$ & 10. & $\tilde{B_y}$ = $\frac{i \left(a C_1 C_2+2 b C_1 C_4+c C_4 C_2\right)}{\left(C_1-C_4\right) \left(C_2^2-C_1 C_4\right)}$ \\
 & $\tilde{W}_2$ =  $ -\frac{i (-a C_4+2 b C_2-c C_1)}{4 C_2 (C_1+C_4)}$& & $\tilde{W}_1$ = $\frac{i \left(a C_1^2-a C_4 C_1+2 a C_2^2+2 b C_2 C_1+2 b C_2 C_4-c C_4 C_1+2 c C_2^2+c C_4^2\right)}{4 \left(C_4 C_1^2-C_4^2 C_1-C_1 C_2^2+C_2^2 C_4\right)}$ \\
\hline
& $\tilde{B_x}$ = 0& &$\tilde{J_3}$ =0 \\
2. &$\tilde{B_y}$ = $\frac{i (a-c)}{2 C_2}$ & 11. &$\tilde{B_y}$ = $-\frac{2 i \left(a C_2 C_1+b C_1^2+b C_4^2+c C_2 C_4\right)}{\left(C_1-C_4\right) \left(C_1^2-2 C_2^2+C_4^2\right)}$ \\
& $\tilde{W}_2$ = $-\frac{i \left(-a C_4+2 b C_2-c C_1\right)}{4 C_2 \left(C_1+C_4\right)}$& &$\tilde{W}_1$ = $-\frac{i \left(-a C_4 C_1-2 a C_2^2+a C_4^2-2 b C_2 C_1-2 b C_2 C_4+c C_1^2-c C_4 C_1-2 c C_2^2\right)}{2 \left(C_1-C_4\right) \left(C_1^2-2 C_2^2+C_4^2\right)}$ \\
\hline
& $\tilde{J_1}$ = 0& &$\tilde{W_3}$ =0\\
3. &$\tilde{B_y}$ = $\frac{i \left(a C_1^2+a C_4 C_1+2 a C_2^2+2 b C_2 C_1-2 b C_2 C_4-c C_4 C_1-2 c C_2^2-c C_4^2\right)}{4 C_2 \left(C_2^2+C_1 C_4\right)}$ & 12. & $\tilde{B}_y$ = $-\frac{i \left(-a C_1+2 b C_2-c C_4\right)}{2 C_2 \left(C_1-C_4\right)}$\\
& $\tilde{W}_2$ = $\frac{i \left(-a C_1^2+a C_4 C_1+2 a C_2^2-2 b C_2 C_1-2 b C_2 C_4+c C_4 C_1+2 c C_2^2-c C_4^2\right)}{8 C_2 \left(C_2^2+C_1 C_4\right)}$& &$\tilde{W}_1$ =$-\frac{i (a+c)}{2 \left(C_1-C_4\right)}$ \\
\hline
& $\tilde{J_2}$ = 0& &$\tilde{B_z}$=0\\
4. &$\tilde{B_y}$ = $\frac{i \left(-a C_1^2-a C_4 C_1+2 a C_2^2-2 b C_2 C_1+2 b C_2 C_4+c C_4 C_1-2 c C_2^2+c C_4^2\right)}{4 C_2 \left(C_2^2-C_1 C_4\right)}$& 13. & $\tilde{W}_1$=$-\frac{i \left(a C_4+2 b C_2+c C_1\right)}{2 \left(C_1-C_4\right) \left(C_1+C_4\right)}$ \\
& $\tilde{W}_2$ = $\frac{i \left(a C_1^2-a C_4 C_1+2 a C_2^2+2 b C_2 C_1+2 b C_2 C_4-c C_4 C_1+2 c C_2^2+c C_4^2\right)}{8 C_2 \left(C_2^2-C_1 C_4\right)}$& &$\tilde{W}_2$=$-\frac{i b}{C_1+C_4}$ \\
\hline
& $\tilde{J_3}$ = 0& &$\tilde{B}_x$ =0\\
5. &$\tilde{B_y}$ = $\frac{i \left(C_1 \left(C_4 (a-c)-2 b C_2\right)+2 C_2^2 (c-a)+a C_4^2+2 b C_2 C_4-c C_1^2\right)}{2 C_2 \left(C_1^2-2 C_2^2+C_4^2\right)}$& 14. &$\tilde{W}_1$=$\frac{i (a-c)}{2 \left(C_1+C_4\right)}$ \\
& $\tilde{W}_2$ = $-\frac{i \left(C_1 \left(C_4 (a+c)+2 b C_2\right)+2 C_2^2 (a+c)-a C_4^2+2 b C_2 C_4-c C_1^2\right)}{4 C_2 \left(C_1^2-2 C_2^2+C_4^2\right)}$& &$\tilde{W}_2$=$-\frac{i \left(-a C_1+2 b C_2-c C_4\right)}{4 C_2 \left(C_1+C_4\right)}$ \\
\hline
& $\tilde{W}_3$ = 0& &$\tilde{J}_1=0$\\
6. &$\tilde{B_y}$ = $-\frac{i \left(a C_4+2 b C_2+c C_1\right)}{2 C_2 \left(C_1-C_4\right)}$& 15. & $\tilde{W}_1$=$\frac{i \left(a C_1^2+a C_4 C_1+2 a C_2^2+2 b C_2 C_1-2 b C_2 C_4-c C_4 C_1-2 c C_2^2-c C_4^2\right)}{4 \left(C_4 C_1^2+C_2^2 C_1+C_4^2 C_1+C_2^2 C_4\right)}$ \\
& $\tilde{W}_2$ = $\frac{i (a+c)}{4 C_2}$& &$\tilde{W}_2$=$\frac{i \left(a C_1 C_2-2 b C_1 C_4+c C_4 C_2\right)}{2 \left(C_1+C_4\right) \left(C_2^2+C_1 C_4\right)}$ \\
\hline
& $\tilde{B}_z$ = 0& &$\tilde{J}_2=0$\\
7.& $\tilde{B}_y$ = $-\frac{2 i b}{C_1-C_4}$& 16. &$\tilde{W}_1$ = $\frac{i \left(a C_1^2+a C_4 C_1-2 a C_2^2+2 b C_2 C_1-2 b C_2 C_4-c C_4 C_1+2 c C_2^2-c C_4^2\right)}{4 \left(C_4 C_1^2+C_4^2 C_1-C_1 C_2^2-C_2^2 C_4\right)}$ \\
& $\tilde{W}_1$ = $-\frac{i \left(a C_4-2 b C_2+c C_1\right)}{2 \left(C_1-C_4\right) \left(C_1+C_4\right)}$& &$\tilde{W}_2$=$\frac{i \left(a C_1 C_2+2 b C_1 C_4+c C_4 C_2\right)}{2 \left(C_1+C_4\right) \left(C_2^2-C_1 C_4\right)}$ \\
\hline
& $\tilde{B}_x$ = 0& &$\tilde{J}_3$ =0\\
8.& $\tilde{B}_y$ = $\frac{i \left(a C_1-2 b C_2+c C_4\right)}{2 C_2 \left(C_1-C_4\right)}$ & 17. &$\tilde{W}_1$=$-\frac{i \left(-a C_1 C_4+2 a C_2^2-a C_4^2+2 b C_1 C_2-2 b C_2 C_4+c C_1^2+c C_1 C_4-2 c C_2^2\right)}{2 (C_1+C_4) \left(C_1^2-2 C_2^2+C_4^2\right)}$\\
& $\tilde{W}_1$ = $-\frac{i \left(a C_4-2 b C_2+c C_1\right)}{2 \left(C_1^2-C_4^2\right)}$& &$\tilde{W}_2$=$-\frac{i \left(a C_1 C_2+b C_1^2+b C_4^2+c C_2 C_4\right)}{(C_1+C_4) \left(C_1^2-2 C_2^2+C_4^2\right)}$ \\
\hline
& $\tilde{J}_1$ = 0& & $\tilde{W}_3$=0\\
9.& $\tilde{B}_y$ = $\frac{i \left(a C_1 C_2-2 b C_1 C_4+c C_4 C_2\right)}{\left(C_1-C_4\right) \left(C_2^2+C_1 C_4\right)}$ & 18. &$\tilde{W}_1$=$-\frac{i \left(a C_4+2 b C_2+c C_1\right)}{2 \left(C_1^2-C_4^2\right)}$\\
& $\tilde{W}_1$ = $\frac{i \left(a C_1^2-a C_4 C_1-2 a C_2^2+2 b C_2 C_1+2 b C_2 C_4-c C_4 C_1-2 c C_2^2+c C_4^2\right)}{4 \left(C_4 C_1^2+C_2^2 C_1-C_4^2 C_1-C_2^2 C_4\right)}$& &$\tilde{W}_2$=$\frac{i \left(a C_1-2 b C_2+c C_4\right)}{4 C_2 \left(C_1+C_4\right)}$ \\
\hline
\end{tabular}
\end{center}
\end{table} 
\end{widetext}

\end{document}